\documentclass[fleqn,10pt]{wlscirep}
\usepackage[utf8]{inputenc}
\usepackage[T1]{fontenc}
\pdfoutput=1 
\title{Broadband Subwavelength Imaging of Flexural Elastic Waves in Flat Phononic Crystal Lenses}

\author[1]{Hrishikesh Danawe}
\author[1,*]{Serife Tol}

\affil[1]{Department of Mechanical Engineering, University of Michigan, Ann Arbor, MI USA 48109}

\affil[*]{stol@umich.edu}

\begin{abstract}
Subwavelength imaging of elastic/acoustic waves using phononic crystals (PCs) is limited to a narrow frequency range via the two existing mechanisms that utilize either the intense Bragg scattering in the first phonon band or negative effective properties (left-handed material) in the second (or higher) phonon band. In the first phonon band, the imaging phenomenon can only exist at frequencies closer to the first Bragg band gap where the equal frequency contours (EFCs) are convex. Whereas, for the left-handed materials, the subwavelength imaging is restricted to a narrow frequency region where wave vectors in PC and background material are close to each other, which is essential for single-point image formation. In this work, we propose a PC lens for broadband subwavelength imaging of flexural waves in plates exploiting the second phonon band and the anisotropy of a PC lattice for the first time. Using a square lattice design with square-shaped EFCs, we enable the group velocity vector to always be perpendicular to the lens interface irrespective of the frequency and incidence angle; thus, resulting in a broadband imaging capability. We numerically and experimentally demonstrate subwavelength imaging using this concept over a significantly broadband frequency range.
\end{abstract}
\begin{document}

\flushbottom
\maketitle
%
%
\thispagestyle{empty}

\section*{Introduction}

Ever since Veselago’s\cite{Veselago1968} discovery of the negative refraction phenomenon in left-hand material (LHM), artificial materials enabling negative refraction of electromagnetic waves have attracted considerable attention. A flat optical super lens with a double negative effective index was demonstrated for the first time by Pendry\cite{Pendry2000}, which broke the diffraction limit resulting in super-resolution imaging of electromagnetic waves. Later, it was found that a flat super lens can also be realized using artificially
engineered periodic structures called photonic crystals\cite{Luo2002,Cubukcu2003,Parimi2003,Berrier2003,Moussa2005}. Luo et al.\cite{Luo2002} first demonstrated the subwavelength imaging of electromagnetic waves in a 2D photonic crystal. Thereafter, the concept of photonic crystals was further extended to acoustic/elastic waves, known as phononic crystals (PCs), which were exploited for negative refraction and subwavelength imaging of acoustic\cite{Yang2004,Zhang2004,Ke2005,Qiu2005,Sukhovich2008,Sukhovich2009,Robillard2011,Veres2012}/elastic\cite{Croenne2011,Dubois2013,DANAWE2022116552} waves. 

\begin{figure}[ht]
\centering
\includegraphics[width=\linewidth]{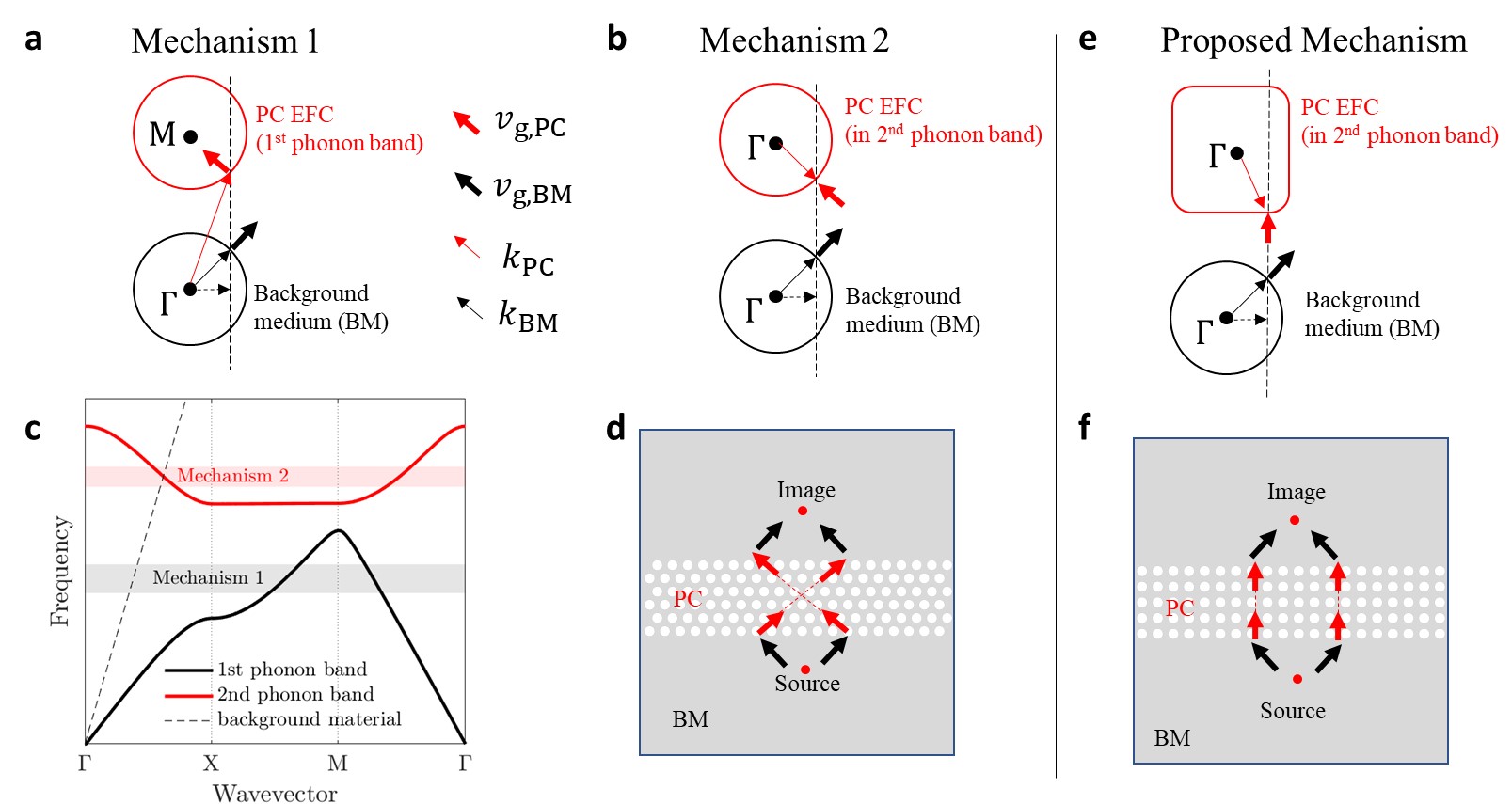}
\caption{\textbf a Imaging mechanism in 1st phonon band of the phononic crystal (PC) near the Bragg band gap where equal frequency contour (EFC) surrounds high symmetry point M. The negative refraction results as the group velocity vector $v_{\textrm{g,PC}}$ is at an angle to wavevector $k_{\textrm{PC}}$ due to convex PC EFC. \textbf b Imaging mechanism in 2nd phonon band due to backward wave propagation. The negative refraction results as the group velocity vector $v_{\textrm{g,PC}}$ is oriented opposite to the wavevector $k_{\textrm{PC}}$. \textbf c The band structure showing the narrow band frequency regions in which imaging mechanism 1 and 2 operate. \textbf d Schematic of negative refraction and subwavelength imaging via mechanism 1 and 2 illustrated using group velocity vectors. \textbf e Proposed imaging mechanism for broadband subwavelength imaging by exploiting flat faces of PC EFC in the 2nd phonon band due to anisotropy. \textbf f Schematic of negative refraction and subwavelength imaging via the proposed mechanism illustrated using group velocity vectors. }
\label{fig:Mechanisms}
\end{figure}

There are two mechanisms for realizing subwavelength imaging in photonic/phononic crystals. The first mechanism utilizes the first phonon band near the Bragg band gap, where intense Bragg scattering results in convex equal frequency contours (EFCs) that enclose high symmetry points of the first Brillouin zone\cite{Luo2002,Cubukcu2003,Zhang2004,Veres2012,Dubois2013}, as depicted in Fig. \ref{fig:Mechanisms}\textbf a. The image formation via the first phonon band was later attributed to the flat-faced EFCs resulting in the self-collimation effect and complex near-field wave scattering effect\cite{Li2003,Kuo2004,He2008}. The second mechanism utilizes opposite directions of the wave vector and energy flow direction at higher (typically the second) phonon band, resulting in negative refraction due to the backward wave effect similar to LHM\cite{Parimi2003,Berrier2003,Moussa2005,Yang2004,Ke2005,Qiu2005,Sukhovich2008,Sukhovich2009,Robillard2011,Croenne2011,DANAWE2022116552}, as depicted in Fig. \ref{fig:Mechanisms}\textbf b. While both mechanisms can be exploited for subwavelength imaging, the first mechanism is advantageous due to the single mode and high transmission, but it only supports near-field images. On the other hand, the second mechanism produces high-quality far-field images via all-angle negative refraction (AANR). However, subwavelength imaging via both mechanisms is limited to a narrow frequency range, as depicted in Fig. \ref{fig:Mechanisms}\textbf c. In the first phonon band, negative refraction only occurs at frequencies close to the band gap where the EFCs are convex. For instance, the 2D photonic crystal proposed by Luo et al.\cite{Luo2002} has convex EFCs in approximately 20\% frequency range about a center frequency near the first Bragg band gap. Later, efforts were made to increase the imaging frequency range in the first phonon band via topology optimization\cite{Yang2017}. The subwavelength imaging in the second band is limited to an even narrower range of frequencies at which the wave vectors in PC and background material have to be close to each other to support AANR. This is essential because of the circular shape of EFCs in the second band of PC requiring equal angles of incidence and refraction (refractive index of -1) for single-point image formation to take place. In the study presented by Robillard et al. \cite{Robillard2011}, the imaging phenomenon is limited to approximately 11\% frequency range about a center frequency. The image formation via mechanism 1 and and 2 is depicted in Fig. \ref{fig:Mechanisms}\textbf d using group velocity vectors. Besides 2D PCs, subwavelength imaging was also studied for Lamb modes in plates via both mechanisms discussed above for the design of flat elastic lenses. However, Lamb modes are more complex than the bulk modes in homogeneous solids, and there are very few studies demonstrating subwavelength imaging of elastic waves in plates via the concept of PCs\cite{Dubois2013,DANAWE2022116552}. Subwavelength imaging was observed for flexural plate mode in the first phonon band by Dubois et al.\cite{Dubois2013}, with the first mechanism utilizing the convex shape of EFCs near the first Bragg bandgap. In our recent work\cite{DANAWE2022116552}, we presented subwavelength imaging of flexural waves in the second phonon band via wave vector matching and the AANR phenomenon. However, all previous lens designs operate in a very narrowband frequency range due to limitations discussed earlier.

In this work, we propose a new methodology for broadband subwavelength imaging of flexural elastic waves in the second phonon band. As opposed to wavevector matching in circular EFCs, our PC design exploits the flat square-shaped EFCs in the second phonon band of a square lattice, as depicted in Fig. \ref{fig:Mechanisms}\textbf e, to achieve subwavelength imaging over most of the second phonon band frequencies.  The flat edges of the EFC enable the group velocity vector in PC always to be perpendicular to the lens interface irrespective of the frequency and incidence angle, as shown in Fig. \ref{fig:Mechanisms}\textbf f. The experimentally validated design shows subwavelength imaging over 50\% frequency range about a center frequency with a resolution as low as 0.63$\lambda$. 

\section*{Results}
\subsection*{Phononic Crystal Design}
\begin{figure}[ht]
\centering
\includegraphics[width=\linewidth]{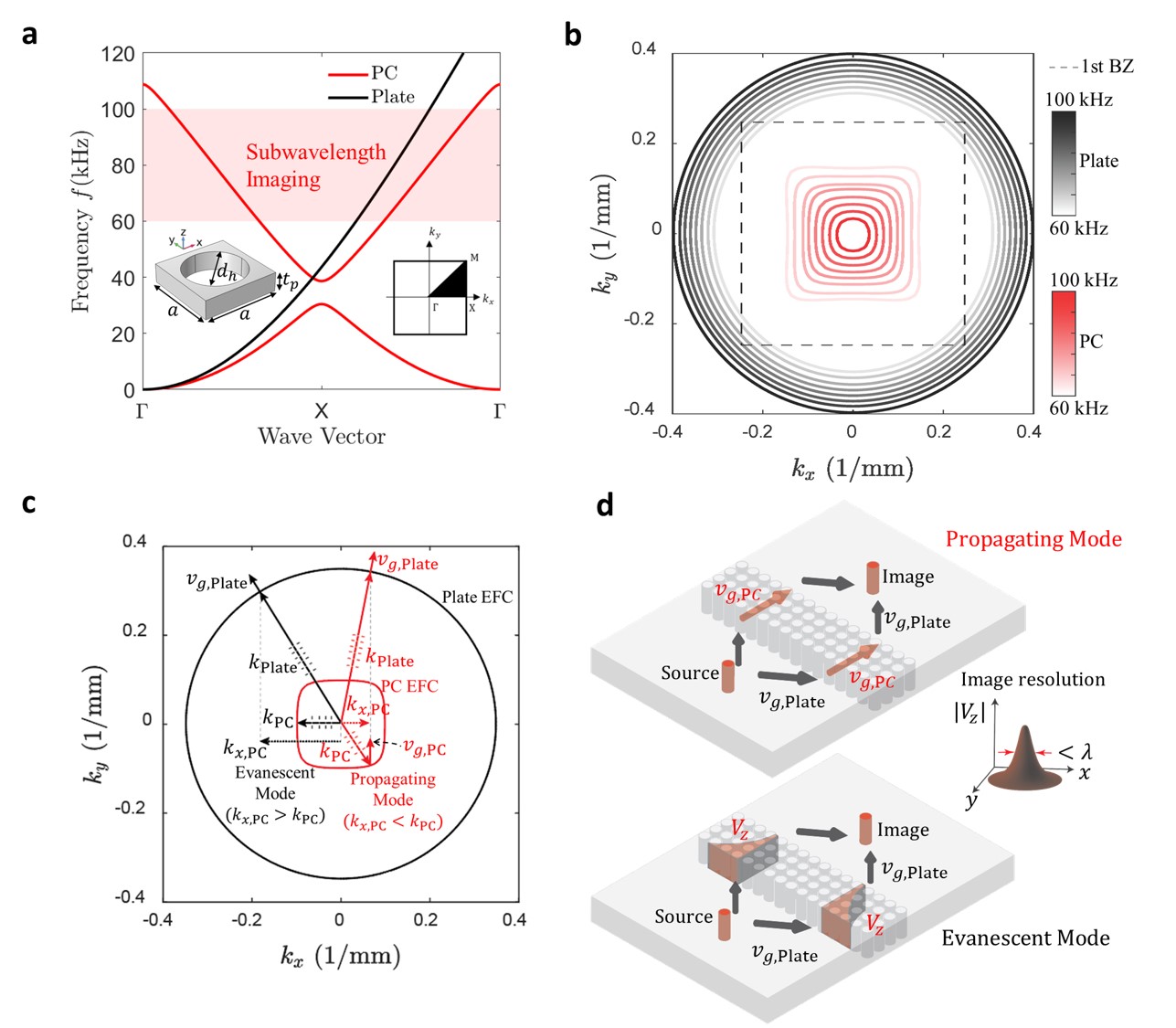}
\caption{\textbf a Band structure of phononic crystal (PC) and plate made of aluminum with the plate thickness $t_p=$3.175 mm. The unit cell of square lattice PC is shown in the inset consisting of an aluminum plate of side length $a=$12.7 mm and a central hole of diameter $d_h=$10.2 mm. \textbf b Equal frequency contours of the aluminum plate and PC in the frequency range of 60 kHz to 100 kHz. \textbf c Wave vectors ($k$) and group velocity vectors ($v_g$) for PC and plate for propagating and evanescent waves. \textbf d Schematic of subwavelength imaging with anisotropic EFC of PC illustrated using group velocity vectors. The group velocity vector in PC ($v_{\textrm{g,PC}}$) remain always perpendicular to the lens interface for the propagating modes and the velocity amplitude ($V_{z}$) decays inside the PC from one edge to other for the evanescent modes.}
\label{fig:Design}
\end{figure}

The dispersion bands of a square lattice PC and aluminum plate are shown in Fig. \ref{fig:Design}\textbf a. The unit cell of PC consists of an aluminum plate with a central hole of a diameter of $d_h=$10.2 mm. The side lengths of the unit cell equal $a = 12.7$ mm, and plate thickness equals $t_p = 0.25a =3.175$ mm. The PC plate is an infinite periodic structure obtained by repeating the unit cell along $x$- and $y$-directions. This geometric periodicity of the structure results in peculiar wave propagation characteristics due to Bragg scattering that creates band gaps in their phonon band structure. The equal frequency contours (EFCs) are 2D projections of the band structure in the wave vector plane at specific frequencies. The EFCs represent wave vectors of plane waves propagating in a different direction in the $x-y$ plane. The EFCs of the PC plate and the baseline uniform aluminum plate of the same thickness corresponding to the flexural (A0) plate mode are depicted in Fig. \ref{fig:Design}\textbf b, where $x$- and $y$-axes of the plot show $x$- and $y$-components of the wave vector. The solid line contours correspond to the EFCs of PC and plate at different frequencies of the second phonon band of PC, and the dotted square contour is the boundary of the 1st Brillouin zone of the PC. More details on band structure calculation can be found in the Methods section.

The EFCs of the square lattice PC form square-shaped contours resulting in flat edges along $x$- and $y$-directions; thus, the PC is anisotropic at the second phonon band frequencies ranging from 60 to 100 kHz. On the other hand, the EFCs of the uniform aluminum plate are perfect circles in the same frequency range representing isotropic behavior. Figure \ref{fig:Design}\textbf c depicts wave vectors and group velocity vectors of the PC and plate for the propagating modes and the evanescent modes of the PC. Since the EFCs of the PC correspond to the second phonon band, the group velocity vector, $v_{g,\textrm{PC}}$ (gradient of the phonon band), points opposite to the wave vector, $k_{\textrm{PC}}$, resulting in backward wave propagation. Whereas; the group velocity vector, $v_{g,\textrm{Plate}}$, of the plate points in the same direction as the wave vector, $k_{\textrm{Plate}}$. Thus, a propagating wave undergoes negative refraction at the interface formed by the PC and plate, as depicted in Fig. \ref{fig:Design}\textbf d, to preserve the energy flow direction ($v_g$) and to satisfy the continuum boundary condition at the interface which requires the wave vector components along the interface (here, ($k_x$)) to be same for PC and plate (i.e., $k_{x,\textrm{Plate}}$=$k_{x,\textrm{PC}}$). Having two such interfaces in the lens enables double negative refraction at the bottom and top edges of the lens. Thus, a point source on one side of the PC lens results in image formation on the other side. The image formation involves two types of modes in the PC lens, i.e., the propagating modes and evanescent modes. The propagating modes occur when the $x$-component ($k_{x,\textrm{Plate}}$=$k_{x,\textrm{PC}}$) of the wave vector of the plate is smaller than the total wave vector in PC ($k_\textrm{PC}$) so that the $y$-component ($k_{y,\textrm{PC}}$) is real by the relation $k_{y, \textrm{PC}}=\sqrt{(k_\textrm{PC}^2-k_{x, \textrm{PC}}^2)}$. On the other hand, if the $x$-component of the wave vector of PC ($k_{x,\textrm{PC}}$) needs to be greater than $k_\textrm{PC}$ to satisfy the continuum condition at the interface, i.e., to match with the $x$-component of plate wave vector ($k_{x,\textrm{Plate}}$), then the $y$-component of PC wave vector ($k_{y,\textrm{PC}}$) becomes imaginary, i.e., $k_{y, \textrm{PC}}=i\sqrt{(k_{x, \textrm{PC}}^2-k_\textrm{PC}^2)}$, resulting in an evanescent wave whose amplitude decays from the bottom edge of the lens towards the top edge. Nevertheless, the propagating modes in PC have group velocity vectors always perpendicular to the lens interface due to the flat edges of the PC-EFCs at all frequencies. Thus, the refracted wave packet at the bottom edge gets transferred to the top edge of the lens, which then undergoes negative refraction for the second time to form an image on the other side of the lens. Hence, flat EFCs in the second phonon band of the anisotropic PC and negative refraction enable broadband subwavelength imaging.

\subsection*{Numerical Results}
\begin{figure}[ht]
\centering
\includegraphics[width=0.9\linewidth]{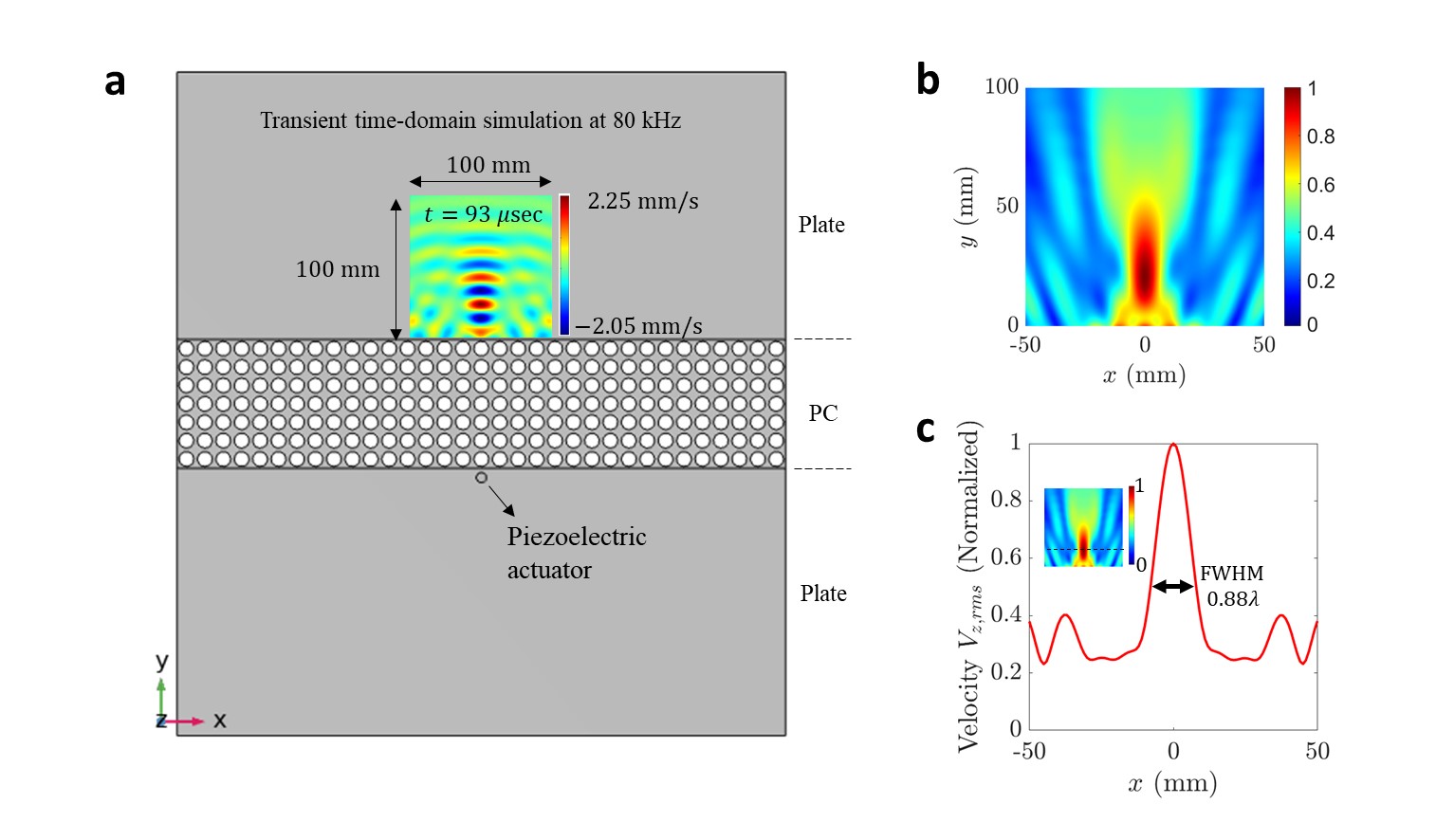}
\caption{\textbf a Finite-element model of PC flat lens for time-domain numerical simulations consisting of 7 x 33 unit cells embedded in an aluminum plate. A thickness mode piezoelectric actuator acts as a source of flexural waves placed at 0.5$a$ distance away from the bottom lens interface. The plate thickness is 0.25$a$ (3.175 mm). The velocity field captured in the image region of 100 x 100 mm at 80 kHz is depicted at time instance 93 $\mu$sec. \textbf b Intensity plot depicting normalized RMS velocity in the image region. \textbf c Normalized RMS velocity line plot along x-direction passing through highest intensity point ($y_{max}$ = 21 mm) in the image region. The intensity map is depicted in the inset. The image resolution is determined using full width at half maximum (FWHM), which is 0.88$\lambda$.}
\label{fig:Sim_Results}
\end{figure}

To test the proposed methodology, we first performed time-domain numerical simulations using a finite element model of an aluminum plate integrated with the PC lens consisting of 7 x 33 unit cells, as depicted in Fig. \ref{fig:Sim_Results}\textbf a. The PC lens is sufficiently long to capture phonon band characteristics of an infinite PC. The dimension of the plate was 419.1 mm x 457.2 mm x 3.175 mm. A piezoelectric actuator disk was used to generate flexural waves in the plate. The actuator was placed 0.5$a$ away from the bottom lens interface, which acts as an omnidirectional source of flexural waves. For more details on simulation settings, please refer to the Methods section.

The flexural waves emanating from the actuator undergo negative refraction at the bottom and top lens interface; thus, focusing wave energy on the other side of the lens. A snapshot of out-of-plane velocity captured for a 4-cycle sine wave signal at 80 kHz is depicted in Fig. \ref{fig:Sim_Results}\textbf a in an image region of 100mm x 100mm size near the top lens interface at a time instance of 93 $\mu$secs. The velocity field shows concentrated regions of high and low wave amplitudes. To obtain the image resolution, normalized root mean square (RMS) velocity representing the intensity map was obtained in the image region, as shown Fig. \ref{fig:Sim_Results}\textbf b. The line plot along the $x$-direction passing through the highest intensity point is depicted in Fig. \ref{fig:Sim_Results}\textbf c. The image resolution is defined based on the full width at half maximum (FWHM) of the line plot along the $x$-direction, which approximately represents the Gaussian distribution of the RMS velocity field. The FWHM at 80 kHz was 0.88$\lambda$, where $\lambda$(=18 mm) is the aluminum plate at 80 kHz. Hence, subwavelength imaging resolution is achieved at 80 kHz. Similarly, simulations were performed at multiple frequencies in the range of 60 kHz to 100 kHz, at which the anisotropic EFCs of PC retain their square shape, and subwavelength imaging was successfully observed at all these frequencies. The numerical results at selected frequencies of 65kHz, 85kHz, and 100 kHz are presented in the next section, along with the experimental results.

\subsection*{Experimental Validation}
\begin{figure}[ht]
\centering
\includegraphics[width=\linewidth]{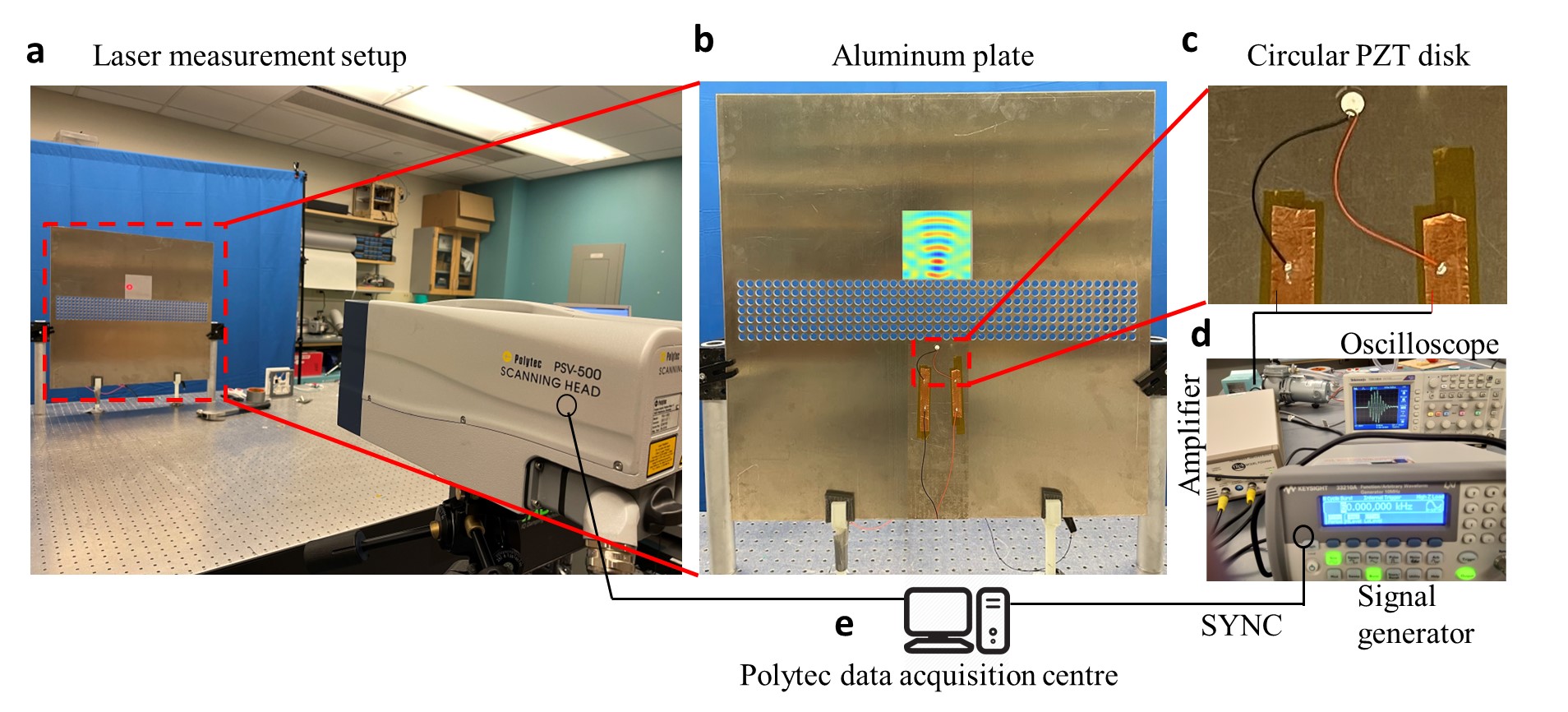}
\caption{\textbf a Experimental setup showing an aluminum plate mounted on a vibration isolation table along with a PSV 500 laser vibrometer for measuring the out-of-plane velocity in the image region. \textbf b Exploded view of the aluminum plate integrated with PC lens consisting of 7 x 45 unit cells. The velocity field captured using a laser vibrometer is depicted in the image region. \textbf c A piezoelectric actuator disk of the subwavelength size is used as a source of omnidirectional flexural waves. \textbf d A signal generator and a power amplifier are used to excite the piezoelectric disk with 4-cycle sine wave signals. An oscilloscope was used to visualize the excited waveform. \textbf e Polytec data acquisition system stored the velocity signal on the plate surface measured by laser vibrometer, and it was in sync with the signal generator.}
\label{fig:Exp_Setup}
\end{figure}

We further validated the subwavelength imaging results through laboratory experiments using a Polytec laser vibrometer and data acquisition system. The experimental setup is depicted in Fig. \ref{fig:Exp_Setup}\textbf a, which shows an aluminum plate clamped using two vertical supports on a vibration isolation table. The plate was further supported at its bottom side using two 3D-printed supports. A Polytec laser vibrometer was used to measure the out-of-plane velocity over the plate surface. An example velocity field capture in the image region is depicted in Fig. \ref{fig:Exp_Setup}\textbf b. The dimensions of the aluminum plate used in the experiments were 609.6 mm x 609.6 mm x 3.175 mm. A 7 x 45 array of holes of diameter 10.2 mm was cut uniformly through the plate using water jetting by maintaining a constant distance between two consecutive holes along $x$- and $y$-directions equaling the lattice constant ($a$). The PC lens size was chosen sufficiently long to capture phononic characteristics of the otherwise infinite periodic structure. A piezoelectric actuator disk of a diameter of 7 mm and thickness of 0.5 mm was used to excite flexural waves in the plate, similar to the numerical models. The actuator was bonded on the plate surface at a distance of 0.5$a$ from the bottom lens interface, as shown in Fig. \ref{fig:Exp_Setup}\textbf c. The piezoelectric actuator was excited using a signal generator connected to a power amplifier depicted in Fig. \ref{fig:Exp_Setup}\textbf d. The signal generator was in sync with laser vibrometer through the Polytec data acquisition system which is schematically represented in Fig. \ref{fig:Exp_Setup}\textbf e. The data acquisition system stored the out-of-plane velocity data measured by the laser on the plate surface. For more details on experimental procedure and data acquisition settings, please refer to the Methods section. 

\begin{figure}[ht]
\centering
\includegraphics[width=\linewidth]{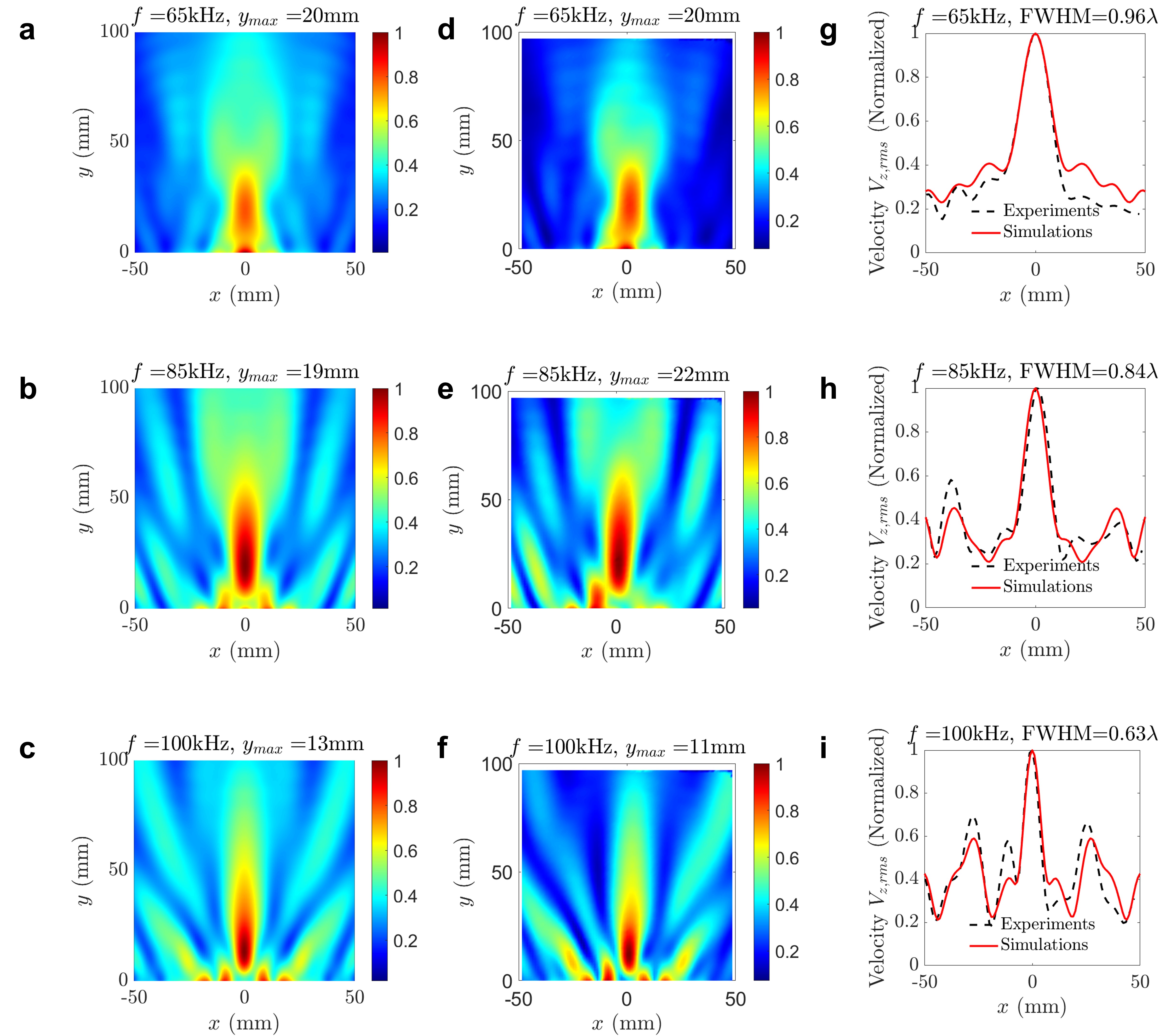}
\caption{\textbf {a, b, and c} Normalized RMS velocity field in the image region obtained from simulations for 65 kHz, 85 kHz, and 100 kHz, respectively. \textbf{d, e, and f} Normalized RMS velocity field in the image region obtained from experiments for 65 kHz, 85 kHz, and 100 kHz, respectively. \textbf{g, h, and i} Comparison of normalized RMS velocity line plots obtained in simulations and experiments along $x$-direction passing through highest intensity point (0, $y_{max}$) for 65 kHz, 85 kHz, and 100 kHz, respectively. The full width at half maximum (FWHM) was obtained for all frequencies in terms of wavelength ($\lambda$). }
\label{fig:Exp_Results}
\end{figure}

The velocity data obtained from laser readings were post-processed to calculate the RMS velocity intensity map at multiple frequencies in the range of 60 kHz to 100 kHz. The measured wavefield at three selected frequencies of 65 kHz, 85 kHz, and 100 kHz are compared with numerical results, as depicted in Fig. \ref{fig:Exp_Results}. The intensity maps obtained from experimental measurements, and numerical simulations at 65 kHz, 85 kHz, and 100 kHz have an excellent agreement. We further compared the experimental and numerical results using the line plots of the intensity map along $x$-direction passing through the highest intensity point in the respective maps. The image resolution was obtained using FWHM of the line plots along the $x$-direction, and it is 0.96$\lambda$, 0.84$\lambda$, and 0.63$\lambda$ at 65 kHz, 85 kHz, and 100 kHz, respectively. We further observed subwavelength imaging resolution at other frequencies in the range of 60 kHz to 100 kHz, with an excellent match between numerical and experimental results. Hence, we have successfully demonstrated broadband subwavelength imaging using flexural waves in a PC aluminum plate.

\section*{Discussion}

In this work, we proposed a PC lens design methodology that achieves subwavelength image resolution in a broadband frequency range as opposed to the previous narrow band designs. We exploit the anisotropy of square lattice that results in flat-face EFCs in second phonon band frequencies. Because of the flat edges of EFCs, the group velocity vector in PC is always perpendicular to the lens interface for all propagating modes, irrespective of the angle of incidence. Thus, all the refracted wave packets at the bottom lens interface get transferred to the top lens interface, and after getting negatively refracted for the second time, they converge back to meet at a single point, forming an image. We numerically and experimentally demonstrated subwavelength imaging of flexural in the frequency range of 60 kHz to 100 kHz, which is 50\% about the center frequency of 80 kHz. The experimental and numerical results are in excellent agreement at all the tested frequencies.

\section*{Methods}

\subsection*{Phonon Band Structure Calculation}
To calculate the phonon band structure of the proposed PC, we utilized the unit cell approach with Floquet periodic boundary conditions (PBC) at unit cell sides along $x$- and $y$-directions representing an infinite PC plate. The Floquet periodic boundary conditions are given as follows:
\begin{equation*}
    \textbf u_{dst}=\textbf u_{src}\cdot e^{i \textbf k \cdot (\textbf r_{dst}-\textbf r_{src})}
\end{equation*}
where $\textbf u_{src}$ and $\textbf u_{dst}$ are displacement vectors at the source and destination boundaries of the unit cell, respectively, similarly, $\textbf r_{src}$ and $\textbf r_{dst}$ are position vectors at the source and destination boundaries of the unit cell, respectively, and $\textbf k$ is the wave vector. The eigenfrequencies of the unit cell with PBC were obtained in COMSOL Multiphysics 5.6 for all wave vectors in the first Brillouin zone. The resulting phononic band structure of flexural mode along $\Gamma X$ symmetry direction of square lattice PC is plotted in Fig. \ref{fig:Design}\textbf a along with the dispersion curve of flexural mode in the aluminum plate. The material properties used for aluminum are $\rho=2700$ kg/m$^{3}$, $E=70$ GPa, $\nu=.33$. The EFC plot in Fig. \ref{fig:Design}\textbf b is obtained using contour plots at respective frequencies.

\subsection*{Finite Element Numerical Simulations}
The time domain numerical simulations were done in COMSOL Multiphysics 5.6. The elastic domain was modeled in the Solid Mechanics module, whereas the piezoelectric effect of the actuator disk was modeled in the Electrostatic module. The plate was assigned free boundary conditions at all its sides. The combined multiphysics simulations were run for 200 $\mu$secs with a fixed time step of 0.1 $\mu$sec. The maximum mesh size was set to $\lambda/10$ using the tetrahedral mesh elements for the plate and the piezoelectric disk. The piezoelectric material was chosen as PZT-5A, and the actuator disk was excited using a 4-cycle sine wave burst with a Gaussian window at the start of the simulation. The simulations were run independently for every frequency from 60 kHz to 100 kHz in steps of 5 kHz. The velocity signal was exported from the image region (100 mm x 100mm) with a spatial resolution of 1mm. The velocity data were post-processed to obtain the RMS velocity field and to determine the image resolution using the intensity map.

\subsection*{Experimental Procedure}

The piezoelectric actuator disk from Steminc Inc. was vacuum bonded using epoxy on the plate surface. The actuator disk vibrates along its thickness direction on applying a voltage signal across its flat faces. The voltage signal was generated using the Keysight 33210A function generator and amplified with a TReK PZD350A amplifier before applying it to the actuator disk. The piezoelectric disk was excited using a 1V (peak-to-peak) 4-cycle sine burst signal with a 50ms delay in between two consecutive bursts. The delay was chosen large enough to let the signal from the previous burst die out completely before sending the next burst so as to avoid any residual signal reflected from plate boundaries interfering with the next burst. The signal generator was in sync with the laser vibrometer through the Polytec data acquisition system. The sampling frequency of laser measurements was set at 6.25 MHz with 2500 sample points. The laser vibrometer was set to measure the out-of-plane velocity on the plate surface by scanning the image region (100mm x 100mm) at every grid point with a spatial resolution of 1.3 mm. A reflective tape was attached to the plate in the image region for clear laser readings. The experiments were run independently for every frequency from 60 kHz to 100 kHz in steps of 5 kHz. The velocity data captured by the laser was post-processed to obtain RMS velocity plots similar to the numerical simulations, and the image resolution was determined from the line plots in the intensity map.

\section*{Data availability}
Data available on request from the authors.

\section*{Acknowledgements}

This work was supported in part by the National Science Foundation [grant number CMMI-1914583].

\section*{Author contributions statement}
H.D. and S.T. conceptualized the idea and developed the methodology, H.D. conducted formal analysis and validation studies, H.D. conducted experiments and analyzed results, H.D. prepared the original draft, S.T. and H.D. reviewed and edited the manuscript. S.T. acquired funding and supervised the study. 

\section*{Competing interests}
The authors declare no competing interests.

\end{document}